\documentclass[twocolumn,prl]{revtex4}

\usepackage{graphicx}
\usepackage{rotating}
\usepackage{amsmath}
\usepackage{amsfonts}
\usepackage{amssymb}
\usepackage{enumerate}
\usepackage{longtable}
\usepackage{dcolumn}
\usepackage{bm}

\begin{document}

\newcommand{\be}{\begin{equation}}
\newcommand{\ee}{\end{equation}}
\newcommand{\bn}{\begin{eqnarray}}
\newcommand{\en}{\end{eqnarray}}

\title{Theory of Normal State Pseudogap Behavior in
FeSe$_{1-x}$Te$_{x}$}

\author{L. Craco$^1$ and M.S. Laad$^2$}

\affiliation{$^1$Max-Planck-Institut f\"ur Chemische Physik fester
Stoffe, 01187 Dresden, Germany \\
$^2$Institut f\"ur Theoretische Physik, RWTH Aachen, 52056 Aachen, Germany}

\date{\rm\today}

\begin{abstract}
The normal state of the recently discovered Iron Selenide (FeSe)-based 
superconductors shows a range of inexplicable features. Along with 
bad-metallic resistivity, characteristic pseudogap features and 
proximity to insulating states, reminiscent of the underdoped high-T$_{c}$ 
cuprates, mark these systems as strongly correlated non-Fermi Liquid 
metals. Here, using the first-principles LDA+DMFT method, we show how 
strong multi-orbital correlation-induced orbital-selective Mott-like 
physics leads to an orthogonality catastrophe underpinning these 
inexplicable incoherent features.  Excellent agreement with a range 
of resistivity and Seebeck data strongly support our proposal. We 
discuss pseudogap regime microscopically, along with implications 
for the nature of the instability at lower $T$, and propose that 
related systems could be of use in thermoelectric devices.   
\end{abstract}


\maketitle

The Iron Selenides (FeSe), with or without Tellurium (Te) substitution,
are the latest addition to a rapidly growing list of Fe-based unconventional
superconductors~\cite{[1]}.  These latter systems are structurally simpler
than their 1111-oxyarsenide counterparts~\cite{[2]};  they nevertheless 
exhibit a host of very unusual physical responses.  These are (by no means 
a complete list): $(i)$ a linear-in-$T$ dependence of the bad metallic 
normal state resistivity, $\rho_{dc}(T)\simeq \rho_{0}+AT$ above $T_{c}$, 
and a Te- and Cu-substitution induced metal-insulator transition (MIT) in
FeSe$_{1-x}$Te$_{x}$~\cite{[3]} and Fe$_{1-x}$Cu$_{x}$Se~\cite{[4]}.
$(ii)$ The NMR relaxation rate is distinctly of the non-Korringa form, and
the static (uniform, ${\bf q}=0$) spin susceptibility, $\chi(T) \simeq T^{1+n}$
with $0<n<1$, in FeSe~\cite{[5]}. $(iii)$ An almost non-existent Drude peak,
with dominant incoherent features in the optical conductivity~\cite{[6]}.
$(iv)$ Incoherent pseudogap like low-energy features in angle-integrated
photoemission (PES)~\cite{[7]}.  Also, an ARPES study for the
antiferromagnetic compositions shows that the AF ordering wave-vector,
${\bf Q_{AF}}$, is very different from that predicted by 
local-density-approximation (LDA)~\cite{[8]}.

In particular, susceptibility and PES data show up the normal state pseudogap
(PG) in FeSe$_{1-x}$Te$_{x}$.  Several of the above features are reminiscent
of the high-$T_{c}$ cuprates in the under- to optimally doped regime, and,
taken together, defy an interpretation in terms of the Landau Fermi Liquid
(LFL). These PG features must manifest themselves in other probes as well.
In fact, recent normal state resistivity and Seebeck data also exhibit PG
features which closely correlate with each other [the PG in $\rho_{dc}(T)$,
as seen in $(d\rho/dT)$ and $(d^{2}\rho/dT^{2})$ near the PG scale 
$T^{*} \equiv T_{PG}$, correlates well with the related anomaly in $S(T)$ 
at the same $T=T^{*}$]~\cite{[9],[10]}.  Further, at low $T$, the 
thermopower does not show the $S(T) \simeq aT$ behavior; the $T$ dependence 
is distinctly {\it slower} than linear just above $T_{c}$. Along with 
$\rho_{dc}(T)\simeq AT$, this confirms that the normal state in 
FeSe$_{1-x}$Te$_{x}$ is {\it not} a LFL metal.  In addition, $S(T)$ at 
high-$T$ flattens out~\cite{[9]}, being well described by a Heikes-like law; 
this is exactly the behavior expected from a Hubbard-like model at high $T$.  
Finally, $S(T,x)$ data~\cite{[9]} reveal an approximate {\it isosbectic} 
point at $T\simeq 100$~K. This is strong evidence for a crossover scale, 
associated with effectively localized (Heikes law in $S(T)$) to incoherent 
metallic conduction, in FeSe$_{1-x}$Te$_x$.

Taken together, the above also imply that the FeSe$_{1-x}$Te$_x$ are strongly
correlated materials proximate to a Mott insulator.  The microscopic
electronic correlation processes underlying the emergence of an incoherent
pseudogapped metallic {\it phase} above $T_{c}~(T_{N})$ in FeSe$_{1-x}$Te$_x$, 
however, await a consistent theoretical understanding.  Like the 1111-systems,
Fe-selenides are multi-orbital (MO) systems with {\it all} five $d$ bands
crossing the Fermi energy ($E_{F}$) in band (LDA) calculations.  Given the
global non-LFL features found experimentally, however, non-trivial extensions
of the LDA to account for dynamical correlation-driven incoherence and 
breakdown of LFL theory are mandatory.  Here, using the state-of-the-art 
LDA+dynamical-mean-field-theory (LDA+DMFT), already used with good 
success~\cite{[11]} for a range of correlated systems, we study this 
pseudogapped incoherent metallic state in FeSe$_{1-x}$Te$_x$ in detail.

In earlier work, we have shown that the incoherent metal~\cite{[our-fese]}
as well as the Te-doping induced metal-insulator transition
(MIT)~\cite{[our-mit]} can be {\it quantitatively} understood using
LDA+DMFT with sizable $d$ band MO correlations.  Here, we
extend these to characterize the PG phase in depth. Specifically,
we show how the Seebeck co-efficient (thermopower), $S(T)$ is also
quantitatively described, and correlate the specific PG anomalies in
both $S(T)$ {\it and} $\rho_{dc}(T)$ with doping, $x$.  Very good
{\it quantitative} accord with extant data reinforces the basic
hypothesis about crucial role of strong correlations.  To our best
knowledge, this is a {\it first} attempt to uncover such features in an
incoherent metal within LDA+DMFT for a {\it real} system.  Finally, we
show how the itinerant-localized duality (Mottness) makes for a high
thermoelectric figure-of-merit $(ZT)$ in such systems at low $T$, and
suggest possibilities for thermoelectric applications using related systems.

Since the basic LDA+DMFT formulation has been used for a variety of real
systems with good quantitative success~\cite{[11]}, we do not present 
it here, but directly describe the electronic transport within LDA+DMFT
scheme.  The Seebeck coefficient, like the conductivity, is {\it exactly}
computable within DMFT using the fully renormalized LDA+DMFT propagators. 
We generalized the DMFT result~\cite{czycholl,[13]} to the five-band 
case relevant for FeSe$_{1-x}$Te$_x$.  This involves the following steps:

We begin with the general expressions for Seebeck and thermal conductivity,
which, respectively, measures the mixed electrical-thermal correlations
$[S(T)]$ and the heat-current correlations $[k(T)]$ at finite $T$:

\be
S(T)=\frac{1}{T} \frac{A_{1}(T)}{A_{0}(T)} \;,
\ee
\be
k(T)=\frac{1}{T} \left(A_{2}(T) -\frac{A_{1}^{2}(T)}{A_{0}(T)} \right) \;,
\ee
with $A_{n}(T)=\frac{e^{2}}{e^{n} \hbar}
\int_{-\infty}^{+\infty} d\omega \phi^{xx}(\omega)
[-f'(\omega)](\omega-\mu)^{n}$ and
\be
\label{eq3}
\phi^{xx}(\omega)=\frac{1}{V}\sum_{\bf k}Tr[v_{x,a}({\bf k})\rho_{a}({\bf k},\omega)v_{x,b}({\bf k})\rho_{b}({\bf k},\omega)] \;.
\ee
Here, $\rho_{a}({\bf k},\omega)=-\frac{1}{\pi} Im \frac{1}{\omega
-\Sigma_{a}(\omega) -\epsilon_{{\bf k},a}}$ is the LDA+DMFT spectral function,
$a=(xy,xz,yz,x^{2}-y^{2},3z^{2}-r^{2})$ represent the five $d$ orbitals,
$\mu$ is the chemical potential, $v_{x,a}({\bf k})$ is the group velocity,
$f(\omega)$ is the Fermi-Dirac function and $V$ the sample volume. In
DMFT, the $A_{n}(T)$ are convertible to integrals over the unperturbed DOS. 
The {\it only} approximation made here is to ignore the ${\bf k}$ dependence
of $v_{x,a}({\bf k})$, i.e, $v_{x,a}({\bf k})\rightarrow v_{x,a}=v$.  In an
incoherent metal, such as we have here, this is justified, since, between
successive hops, a carrier in an incoherent state does not exist long enough
in a given ${\bf k}$ eigenstate.  Another related reason is the bad-metallic
resistivity, implying $k_{F}l\simeq O(1)$ and leading to the same conclusion.
With this simplification, Eq.~(\ref{eq3}) for the intra-band contribution reads
$\phi^{xx}(\omega)=\frac{v^{2}}{V}\sum_{a \bf k} \rho_{a}^{2} ({\bf k},\omega)$.

\begin{figure}[thb]
\includegraphics[width=\columnwidth]{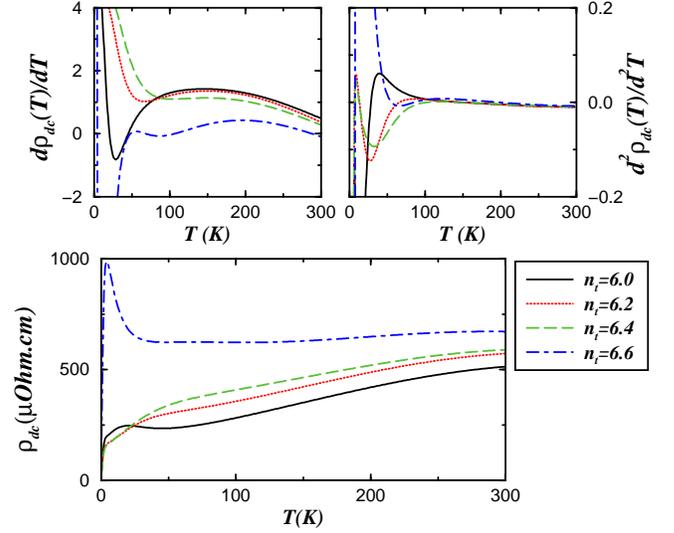}
\caption{(Color online)
Resistivity versus temperature (lower panel) and its derivatives (upper panel)
showing their evolution with increasing $x$ ($n_t=6+x$) within LDA+DMFT for
fixed $U=4.0$~eV and $J_H=0.7$~eV. The pseudogap scale is
$T_{PG}\simeq 150$~K (see text).}
\label{fig1}
\end{figure}

We now describe our results. In Fig.~\ref{fig1}, we show the resistivity,
$\rho_{dc}(T) \equiv 1/A_0(T)$ in FeSe$_{1-x}$Te$_{x}$, derived
earlier~\cite{[our-mit]}. It clearly shows the ``S-like'' shape
characteristic of a PG metal.  At low $T$, $\rho_{dc}(T)\simeq AT$ and
the LFL-like $T^{2}$ form is {\it never} observed.  With increasing $x$,
$\rho_{dc}$ becomes more bad-metallic, smoothly going over to an
insulator-like form up to very low $T\simeq 2.0$~K.  In reality, at much 
higher $T$, either antiferromagnetic (AF) or unconventional superconducting 
(USC) transition will cut-off this extremely low crossover to a metal. For 
non AF/USC ordered systems, even a minute amount of non-magnetic disorder 
on FeSe layers (e.g, Cu~\cite{[4]} will immediately destroy metallicity, 
leading to a metal-insulator transition, as indeed seen in 
Fe$_{1-x}$Cu$_{x}$Se for $x<<1$. Focussing on $x=0$ to make contact with 
experiment~\cite{[10]}, we show 
$(d\rho_{dc}/dT)$ and $(d^{2}\rho_{dc}/dT^{2})$ in the top panels of
Fig.~\ref{fig1}. The PG scale, $T^{*}$ is the temperature at which
$(d\rho_{dc}/dT)$ shows a maximum, while $(d^{2}\rho_{dc}/dT^{2})$ crosses
zero~\cite{[10]}.  The Seebeck co-efficient, $S(T)$, should also show an 
anomalous feature, i.e, a minimum, around $T^{*}$~\cite{[9],[10]}. 

\begin{figure}[thb]
\includegraphics[width=\columnwidth]{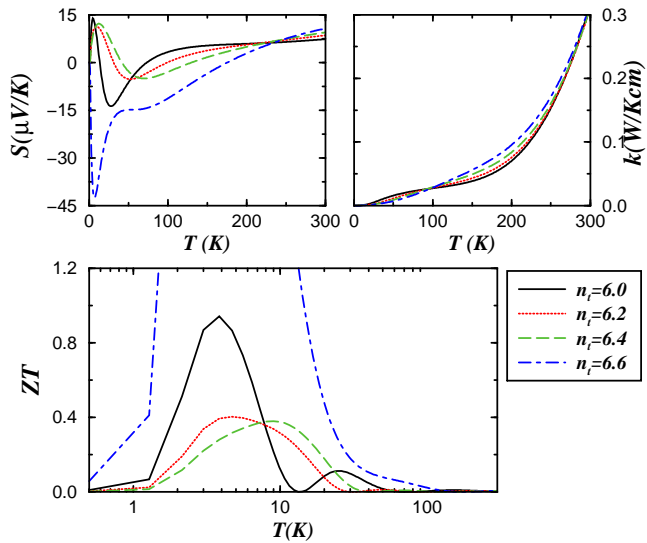}
\caption{(Color online) Seebeck coefficient (top left) and thermal
conductivity (top right) for FeSe$_{1-x}$Te$_{x}$ within LDA+DMFT. Notice
the $S(T)=$const. form at high-$T$, decreasing around $T_{PG}$ and showing
a second crossing at much lower $T_{an}$, all in excellent agreement
with data~\cite{[9],[10]}. The thermoelectric figure of merit, $ZT$, is 
shown in the lower panel. Notice the high value of $ZT$ for $x=0.6$.}
\label{fig2}
\end{figure}

Very good {\it quantitative} agreement with extant data~\cite{[9],[10]} 
is also clearly visible in $S(T,x)$, Fig.~\ref{fig2}.  Our results are 
obviously valid only for $T>T_{c}$, and, at very low $T$, correspond to 
what would be observed in a non-SC system in absence of disorder. A 
number of characteristic features, intimately correlated with those in 
the resistivity derivatives, are noteworthy: $(i)$ $S(T)$ is 
approximately $T$-independent at $T>T_{min}(x)$, in nice 
accord with Heikes law.  Here, it arises due to thermal transport
involving Mott localized carriers in our five-orbital Hubbard model.
$(ii$) In the relevant range of $0\leq x\leq 0.6$, we find that $S(T)$
smoothly decreases, going through a minimum at another scale $T^{*}(x)$.
Interestingly, the deviation from Heikes law sets in around $T_{PG}=150$~K. 
Further, this scale coincides with the PG scale extracted from resistivity
derivatives (see Fig.~\ref{fig1}), pinning down the temperature below which a
low energy PG opens up in transport. $(iii)$ $S(T,x)$ crosses zero {\it twice}
at $T_{low}(x),T_{high}(x)$ (labelled $T_{an}$, $T^{*}$ in Ref.~\cite{[10]}),
again in precise accord with data. $(iv)$ Even at low $T\simeq 10$~K, $S(T)$
does not recover the linear-in-$T$ form: an examination shows a $T^{\eta}$
law with $\eta\leq 1$, suggesting that collective {\it bosonic} fluctuations
dominate the low-$T$ thermopower. $(v)$ Also the approximate {\it isosbectic}
point in $S(T,x)$ as $x$ is varied is reproduced by LDA+DMFT. $(vi)$
Interestingly, $S(T)$ only shows two sign changes for metallic (above $T_{c}$)
compositions.  For large $x>0.5$, appearance of insulator-like behavior
{\it above} $T_{c}$~\cite{[our-mit]} goes hand-in-hand with only one sign
change at high $T^{*}\simeq 150$~K in $S(T)$, also consistent with
findings~\cite{[9]}.

\begin{figure}[thb]
\includegraphics[width=\columnwidth]{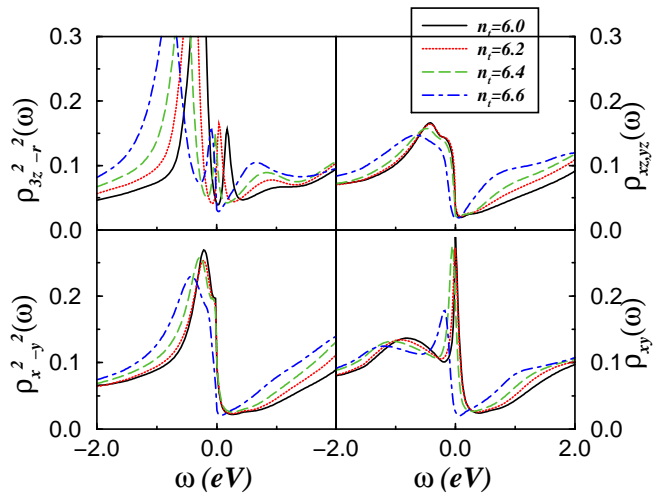}
\caption{(Color online) LDA+DMFT spectral functions for FeSe$_{1-x}$Te$_{x}$.
Clear pseudogap stabilization with increasing $x$ going hand-in-hand with
large scale SWT is apparent.  Power-law fall-offs in the $d_{xz,yz,x^{2}-y^{2}}$
spectra above the low-energy PG, associated with orbital-selective incoherence,
are also visible.}
\label{fig3}
\end{figure}

Remarkably, the {\it full} set of our theoretical $\rho_{dc}(T,x)$ {\it and}
$S(T,x)$ results are in very good agreement with a range of extant results
obtained by two groups~\cite{[9],[10]}.  Our work is the {\it first} 
detailed theoretical study of electrical {\it and} thermal transport in 
Fe-based superconductors. These remarkable results clearly call for a 
deeper microscopic rationalization. Since vertex corrections to 
conductivities rigorously vanish in DMFT~\cite{[16]}, these features 
must be intimately linked to the detailed evolution of the LDA+DMFT 
spectral functions with $T$ and $x$.  In Fig.~\ref{fig3}, we show the 
LDA+DMFT orbital-reolved DOS, $\rho_{a}(\omega)$, for the doping values 
above. Clear signatures of normal state incoherent
metal behavior, along with a low-energy PG, are visible in the DOS,
providing a clear link between PG features in transport and spectral
responses. In particular, we observe important changes in the LDA+DMFT
DOS with $x$, and these must be correlated with features in transport. 
The $d_{xz,yz}$ DOS {\it always} has a deep PG, while the $d_{xy}$ DOS
rapidly loses the small LFL coherence with $x$.  The $d_{x^{2}-y^{2}}$ DOS
shows a progressively deeper PG feature with $x$, while the reverse 
occurs for the $d_{3z^{2}-r^{2}}$ DOS.  These go hand-in-hand with large
changes in spectral weight transfer (SWT), a characteristic feature of
Mottness~\cite{[phillips]}.  Further, clear power-law fall-off of the
DOS at energies above the (small) PG feature around $E_{F}$ instead of a
LFL peak at $E_{F}$ is seen for the $d_{xz,yz,x^{2}-y^{2}}$ DOS.  Finally,
the low-energ kink feature clearly seen for $x=0$ (solid curve) is
smoothened with $x$.  Based on the agreement with transport above, we
predict that PES (ARPES) measurements on FeSe$_{1-x}$Te$_{x}$ as a
function of $x$ should show {\it all} these features in the spectral
function with $x$.

Also the anomalies in $S(T)$ and $(d^{n}\rho_{dc}/dT^{n})$ with $n=1,2$ at 
the much lower $T_{an}$~\cite{note}, observed for the {\it first} time in 
a correlated system~\cite{[10]}, are recovered too within our LDA+DMFT. 
In a band-LFL picture, this behavior would be associated with {\it hole} 
conduction, but such a connection is tenuous in the incoherent metal we 
find.  In real FeSe$_{1-x}$Te$_x$, USC and/or AF actually cuts off this 
behavior, leading to $S(T\rightarrow 0)=0$.  Looking closer at the 
$T\rightarrow 0$ limit within our uniform, symmetry-unbroken, state 
calculation, we find that $S(T\rightarrow 0)\simeq T^{\eta}$ with 
$\eta\leq 1$. Correspondingly, $\rho_{dc}(T)\simeq AT$ up to lowest $T$. 
Neither $\rho_{dc}(T)\simeq T$ nor $S(T)\simeq T^{\eta}$ with $\eta\leq 1$ 
at low $T$ are interpretable in terms of transport associated with LFL 
quasiparticles; rather, they suggest transport in terms of incoherent, 
collective excitations in the non-LFL metal.  {\it If} the above features 
are related to the mechanism of superconductivity (SC)~\cite{[10]}, our 
analysis has far-reaching implications for the SC-instability. Specifically, 
our results now imply that SC must arise directly as an instability of a 
normal non-LFL metal without long-lived fermionic quasiparticles, and 
that the $d_{xz,yz,x^{2}-y^{2}}$ bands will play a central role in pairing, 
with the rest playing a secondary role, as in a proximity effect scenario.  
Recall that, in our earlier LDA+DMFT study of 
FeSe~\cite{[our-fese],[our-mit]}, the non-LFL metal arises from orbital
selective incoherence having its origin in the interplay between strongly
orbital-dependent hopping (as in LDA) and strong dynamical inter-orbital
correlations (DMFT); the non-LFL features are assigned to the generation
of an Anderson {\it orthogonality catastrophe} in the local problem of
DMFT~\cite{[our-fese]}. Once this occurs, the spectral functions will exhibit
a low energy PG with power-law like tails at higher energy, precisely as we
find in the DMFT. In such a metal, the transport properties are dominated
by collective, multi-particle {\it bosonic} excitations~\cite{[hidden-FL]},
rather than by LFL quasiparticles. Our LDA+DMFT results are consistent with 
this incoherence scenario based on Mottness, and excellent agreement with 
data from a number of groups, as shown above, strongly supports our 
contention.

Therefore, we extend our analysis to propose that related
materials have the potentiality to be of use as good thermoelectrics, at
least at low $T$.  To this end, we also need the thermal conductivity,
$k(T)$, also computable {\it exactly} in DMFT in terms of the LDA+DMFT
propagators.  In Fig.~\ref{fig2}, we show the electronic part of the
thermal conductivity $[k(T)]$, along with the thermoelectric figure of
merit, $ZT=\frac{T S^{2} \sigma}{k}$.  We find $k(T)\simeq T^{2}$ at low $T$,
indicating that {\it bosonic} contributions dominate the electronic
contribution to $k(T)$.  We suggest that measuring $k(T)$ at low $T$ will
bare this contribution.  At higher $T>10$~K, additional phononic
contributions will increase $k(T)$ and decrease the electrical conductivity
$\sigma(T)\equiv A_0(T)$, reducing $ZT$. Thus, given the weak electron-phonon
coupling in the Fe-based SC~\cite{[boeri]}, we expect a small-to-moderate
reduction in $ZT$ at low $T$.  Nevertheless, the high value of $ZT$, albeit
at low $T$, is encouraging, and ways of improving $ZT$ at higher $T$, based
on appropriate engineering are an attractive possibility. In particular,
consideration of more geometrically frustrated members of the family of
Fe arsenides and selenides could yield an appreciable $S(T)\simeq T$ at
high $T$ arising from large spin and orbital fluctuations, as in
NaCo$_{2}$O$_{4}$~\cite{[nacoo124]}, and increasing $ZT$.

Given an incoherent non-LFL metal with low energy PG as above, the low $T$
instabilities cannot involve usual particle-hole (p-h) (AF) or
particle-particle (p-p) (USC) pairings in a LFL, since there are {\it no}
coherent LFL quasiparticles in the non-LFL state in the first place. 
We believe that these instabilities will then arise by studying the dominant
two-particle instabilities of such a metal.  As in coupled $D=1$ Luttinger
liquids~\cite{[pwa]}, in a regime where one-particle inter-orbital mixing
is irrelevant, generation of {\it effective} two-particle (pair hopping)
terms in both p-h and p-p sectors at two-loop level in a renormalization
group treatment around the {\it impurity} (DMFT) limit lead to ordered states
emerging directly from the incoherent metal~\cite{[our-feas-SC]}.  We leave
the details of the mechanism of USC from an incoherent metal derived above
to the future.

In conclusion, we have studied the detailed nature of the incoherent metal
with a low energy pseudogap using the first-principles LDA+DMFT. The anomalous
character of the transport properties is interpreted in terms of electrical
and heat transport by {\it collective}, multi-particle excitations.  These
arise from an Anderson orthogonality catastrophe having its origin in the
selective-Mott localization in LDA+DMFT. Excellent semiquantitative agreement
with a whole host of transport ($\rho_{dc}(T)$ and $S(T)$) data with $x$
puts our mechanism on solid ground.  Our study puts strong theoretical
constraints on how the ordered states (AF/USC) arise directly from the
incoherent metal, and shows the crucial importance of Mottness in this
context.


\begin{thebibliography}{28}
\bibitem{note} $T_{an}$ is the temperature where $d^{2}\rho_{dc}/dT^{2}$
changes sign at low-$T$.
\end{thebibliography}
\end{document}